\documentclass[article]{IEEEtran}
\usepackage{color}
\usepackage{needspace}

\input{mysymbol.sty}
\usepackage{theorem}
\usepackage{cite}
\usepackage{amsmath}
\usepackage{amssymb}
\usepackage{mathtools}
\usepackage{multirow}
\usepackage{booktabs}

\usepackage{url}
\usepackage{graphics, subfigure, times, amsfonts}
\usepackage{tikz, epic,eepic}
\usetikzlibrary{shapes,arrows}
\usepackage{pgfplots}
\usepackage{color}
\usepackage{hyperref}

\usepackage{srcltx} 
\usepackage{latexsym}
\usepackage{amscd, verbatim}

\selectfont

\newtheorem{theorem}{\hspace{0pt}\bf Theorem}
\newtheorem{assumption}[theorem]{\hspace{0pt}\bf Assumption \hspace{-0.15cm}}
\newtheorem{condition}[theorem]{\hspace{0pt}\bf Condition \hspace{-0.15cm}}
\newtheorem{lemma}[theorem]{\hspace{0pt}\bf Lemma}

\newtheorem{algorithm}[theorem]{\hspace{0pt}\bf Algorithm}

\newtheorem{remark}[theorem]{\hspace{0pt}\bf Remark}

\newcommand{\bru}{\hhatu}
\newcommand{\brsigma}{\hat{\sigma}}
\newcommand{\ones}{{\bf{1}}}

\newcommand{\R}{\mathbb{R}}
\newcommand{\N}{\mathcal{N}}

\def\forall{\text{for all\ }}
\def\N{\mathbb{N}}

\mathtoolsset{showonlyrefs}

\title{Distributed Inertial Best-Response Dynamics}

\author{Brian Swenson$^{1}$, Ceyhun Eksin$^{2}$,  Soummya Kar$^{1}$, Alejandro Ribeiro$^{3}$
\thanks{*{The work of B. Swenson and S. Kar was supported in part by NSF grant CCF-1513936.}}
\thanks{$^{1}$Dept. of Electrical and Computer Engineering, Carnegie Mellon University, Pittsburgh, PA.  {\tt\small brianswe@ece.cmu.edu, soummyak@andrew.cmu.edu}}
\thanks{$^{2}$Industrial \& Systems Engineering Department, Texas A\&M University, College Station, TX 77843.
        {\tt\small ceyhuneksin@gatech.edu}}%
\thanks{$^{4}$Dept. of Electrical and Systems Engineering, University of Pennsylvania, Philadelphia, PA.  {\tt\small aribeiro@seas.upenn.edu}}%
}

\begin{document}
\normalsize
\maketitle

%
\begin{abstract}
The note considers the problem of computing pure Nash equilibrium (NE) strategies in distributed (i.e., network-based) settings. The paper studies a class of inertial best response dynamics based on the fictitious play (FP) algorithm. It is shown that inertial best response dynamics are robust to informational limitations common in distributed settings. Fully distributed variants of FP with inertia and joint strategy FP with inertia are developed and convergence is proven to the set of pure NE. The distributed algorithms rely on consensus methods. Results are validated using numerical simulations.
\end{abstract}

\section{Introduction}\label{sec_intro}
In this note we are concerned with the problem of distributed computation of pure-strategy Nash equilibria (NE) in finite games.
More precisely, we are interested in a scenario in which a group of agents, capable of communicating over a sparse communication network, would like to cooperatively compute a Nash equilibrium of some associated game.

As an example, consider the problem of distributed UAV target assignment \cite{Arslan_et_al_2007}. Suppose a group of UAVs is tasked with covering a set of targets---each target should be covered by (or assigned to) at least one UAV. The UAVs are capable of communicating with neighboring UAVs using a short range radio. It is desired that, using the ad-hoc communication network, the UAVs negotiate on an acceptable target assignment which they can then physically implement. The target assignment problem can be modeled as a game, the equilibria of which are acceptable target assignments. The problem thus reduces to one of distributed computation of Nash equilibria prior to physically engaging in some game.\footnote{The work \cite{Arslan_et_al_2007} considers a similar target assignment problem, but does not consider distributed (i.e., network-based) algorithms for addressing these problems.}


A popular method for computing NE in games is the use of so-called game-theoretic learning algorithms, in which players repeatedly play some game, adapting their strategy in each round according to some predefined behavior rules \cite{Fudenberg_Levine_1998,Young_2004}. A particularly simple and useful class of algorithms are those based on best-response adaptation. In such algorithms, players track some statistic of the game (e.g., the empirical distribution of play, or some other useful aggregate statistic) and use this information to forecast how other players will behave in the future. Players choose next-stage actions as a best-response given their forecast.
A Nash equilibrium is, by definition, the fixed point of the best response correspondence and best-response based dynamics play a fundamental role in the field game-theoretic learning \cite{Fudenberg_Levine_1998,Young_2004}.
The class of algorithms based on best-response adaptation is broad, including simple round robin best-response dynamics \cite{nisan2007algorithmic}, fictitious play (FP) \cite{Brown_1951}, and inertial best response dynamics such as FP and joint strategy fictitious play (JSFP) with inertia \cite{Marden_et_al_2009}.

In general (when players are provided with full information about the history of game play) such dynamics are guaranteed to converge to the set of NE in many games of interest, including the class of weakly acyclic games \cite{Young_2004}. While the set of NE includes both mixed (probabilistic) and pure (deterministic) equilibria, in many applications of interest, pure equilibria are preferable to mixed (e.g., in the target assignment problem considered earlier).
The incorporation of an inertial component in best response dynamics is a common technique used to ensure convergence to pure-strategy equilibria. Such dynamics are popular in practice, with a prominent example being joint strategy fictitious play (JSFP) with inertia \cite{Marden_et_al_2009}.


The main contribution of this note is the development of algorithms for computing pure-strategy NE in a distributed setting. In particular, we develop techniques for implementing inertial best-response algorithms in a distributed setting. Our main contributions are the following: (1) We show that inertial best response dynamics are robust to certain types of informational limitations common in distributed settings, (2) We develop a distributed variant of FP with inertia and prove convergence to pure NE in the class of weakly acyclic games; and (3) We develop a distributed implementation of JSFP with inertia and show convergence to pure NE in congestion games. While congestion games constitute a narrower class of games than weakly acyclic games, the informational overhead associated with JSFP is significantly less than that of FP.

We briefly review recent related literature on distributed game-theoretic learning algorithms.
The work \cite{swenson2012distributed} studies a network-based variant of FP for computing NE, \cite{Koshal_et_al_2012} studies a gossip-based algorithm for computing NE in aggregative games, \cite{chen2012spatial} studies  an  algorithm  for  finding  NE  in  a  spatial  spectrum  access  games,  \cite{gharesifard2013distributed}  studies  a  network-based  algorithm  for  NE  seeking  in  a  two-network  zero-sum  games,  \cite{Li_Marden}  presents  a  method  for designing games with a prescribed local dependence, \cite{gharehshiran2013distributed} studies a distributed regret-based reinforcement learning algorithm for tracking the polytope of correlated equilibria in time-varying  games,  and  \cite{salehisadaghiani2016distributed} studies  a  gossip-based  algorithm  for  computing  NE  in  a  network-based setting in games with continuous-action spaces. To the best of our knowledge, the present work is the first to consider the problem of distributed computation of pure strategy NE in finite games.


The remainder of the paper is organized as follows. Section \ref{sec_prelims} sets up notation, Section \ref{sec_repeated_game_play} presents inertial best response dynamics and proves a basic robustness result, Section \ref{sec_D-FP} presents distributed FP with inertia, Section \ref{sec_D-JSFP} presents distributed JSFP with inertia, Section \ref{sec_simulation} gives a simulation example, and Section \ref{conclusion} concludes the paper.


%
\section{Preliminaries} \label{sec_prelims}
A game in normal form is represented by the tuple $\Gamma := (\calN,(\calA_i,u_i)_{i\in \calN})$, where $\calN = \{1,\ldots,n\}$ denotes the set of players, $\calA_i$ denotes the finite set of actions available to player $i$, and $u_i:\prod_{i\in \calN}\calA_i \rightarrow \mathbb{R}$ denotes the utility function of player $i$. Denote by $\calA:= \prod_{i\in \calN} \calA_i$ the joint action space.

We suppose players are permitted to use probabilistic strategies. Formally, let the \emph{mixed strategy space} of player $i$ be given by the set $\triangle(\calA_i)$ of probability distributions over $\calA_i$, and let $\triangle^n(\calA):=\prod_{i=1}^n \triangle(\calA_i)$ denote the set of joint mixed strategies where it is assumed that players use independent strategies. We represent a joint mixed strategy $\sigma\in\triangle^n(\calA)$ as the $n$-tuple $\sigma = (\sigma_1,\ldots,\sigma_n)$, where $\sigma_i\in\triangle(\calA_i)$ denotes the marginal mixed strategy of player $i$. When a mixed strategy $\sigma\in\triangle^n(\calA)$ is played, we are interested in the expected payoff which, in a slight abuse of notation, we write as
\begin{equation}
\label{def_mixed_U}
u_i(\sigma) := \sum_{a \in \calA} u_i(a)\sigma_1(a_1)\ldots \sigma_n(a_n)
\end{equation}
\vskip-5pt
The notation $u_i(\sigma_i,\sigma_{-i})$ is meant to emphasize that the payoff depends on the strategy $\sigma_i$ chosen by player $i$ and the strategies $\sigma_{-i}:=(\sigma_j)_{j\in\calN\backslash\{i\}}$ that are chosen by other players.

Given a strategy $\sigma_{-i}\in \prod_{j\in\calN\backslash\{i\}} \triangle(\calA_i)$, the \emph{best response} set of player $i$ is given by $BR_i(\sigma_{-i}):=\arg\max_{\sigma_i\in\triangle(\calA_i)} u_i(\sigma_i,\sigma_{-i})$.
A mixed strategy $\sigma\in\triangle^n(\calA)$ is said to be a \emph{Nash equilibrium} if $\sigma_i \in BR(\sigma_{-i})$ for all $i\in\calN$.
An equilibrium $\sigma$ is said to be a \emph{pure Nash equilibrium} if there exists an action tuple $a$ such that $\sigma$ places weight 1 on $a$.

The learning algorithms considered in this paper assume the following format of repeated play. Let a normal form game $\Gamma$ be fixed. Let players repeatedly face off in the game $\Gamma$, and for $t\in\{1,2,\ldots\}$, let $a_{i,t}\in \calA_i$ denote the action played by player $i$ in round $t$. Let the $n$-tuple $a_t = (a_{1,t},\ldots,a_{n,t})$ denote the joint action at time $t$.

In this note we will be interested in algorithms for computing NE in a distributed information setup. We will say an algorithm is distributed if it satisfies the following assumption.
\begin{assumption} \label{a_comm_graph}
Players are equipped with a pre-assigned, possibly sparse, communication graph $G=(V,\mathcal{E})$, in which a vertex represents a player, and an edge from vertex $i$ to $j$ represents the ability of player $i$ to communicate information to player $j$. The directed graph $G$ is strongly connected. Players may exchange information with immediate neighbors (the set of neighbors of a player $i$ is given by $\calN_i := \{j\in\calN:(i,j) \in \calE\}$) once between iterations of the repeated play process. Players know the structure of (only) their own utility function, and may directly observe (only) their own actions.
\end{assumption}
We emphasize that players do not know the utility functions of others, nor can they observe the actions of others, nor measure their received payoffs (consider, for example, the distributed target assignment problem in Section \ref{sec_intro}). All additional information must be disseminated over the communication graph.

We remark that in this paper we do not study communication as a strategic element. We are interested in cooperative computation of NE in mutli-agent settings and we assume that agents communicate as prescribed by the associated distributed algorithm.

%

\section{Best-Response Dynamics} \label{sec_repeated_game_play}

Suppose that players are engaged in repeated play of some game $\Gamma$.
For each $i\in\calN$, let $\sigma_{i,t}\in\triangle(\calA_i)$ denote the strategy used by player $i$ in round $t$.
Suppose that prior to round $t$, each agent $i$ forms an estimate $\hat \sigma_{-i,t}^i\in \prod_{j\in\calN\backslash\{i\}} \calA_j$ of the mixed strategy
that will be used by other agents in the upcoming round. The estimated strategy $\hat \sigma_{-i,t}^i$ allows agent $i$ to estimate the payoff that it would receive from playing an arbitrary action $a_i\in\calA_i$. These estimated payoffs can be written as
\begin{equation} \label{eqn_estimated_utility}
   \bru_{i,t}(a_i):= u(a_i, \brsigma_{-i,t}^{i}),~~ a_i\in \calA_i.
\end{equation}
In a best response learning algorithm, in each stage of the repeated play, each player plays an action that maximizes her utility given her estimate of the strategies of others.

In this paper, we are interested in a slight modification in which agents are sometimes ``reluctant'' to modify their action choices from round to round. We refer to this general algorithm (formally stated next) as \emph{inertial best response dynamics}.\footnote{We also refer to an algorithm of this form as an inertial best response algorithm.}

%
\begin{algorithm}\label{algo_fp_inertia} \normalfont
Let $\rho\in (0,1)$ be an inertia constant and let $a_{i,1}$ be an arbitrary initial action for each $i$. At time $t>1$, agent $i$ has access to strategy estimates $\brsigma_{j,t}^{i}$ that it uses to compute the best response set $BR(\hat \sigma_{-i,t}^i)$.
Players are said to follow inertial best response dynamics if they play actions according to
\begin{align}\label{eqn_def_inertia}
   \mathbb{P}\left(a_{i,t+1} = a_{i,t}\vert \mathcal{F}_{t-1}\right) & = \rho,\\
   \mathbb{P}\left(a_{i,t+1} \in BR_i(\hat \sigma_{-i,t})\vert \mathcal{F}_{t-1}\right) & = 1 - \rho,
\end{align}
\end{algorithm}
where $(\mathcal{F}_t)_{t\geq 1}$ is a filtration (sequence of increasing $\sigma$-algebras) that contains the information available to players in round $t$.
%
As per \eqref{eqn_def_inertia}, an inertial best response algorithm entails player $i$ sticking to its previous play $a_{i,t}$ with some (fixed) probability $\rho$ (this is the inertia component of the algorithm), and playing a best response otherwise.


\subsection{General Assumptions}
Unless otherwise specified, throughout the paper we will consider inertial best response dynamics in games satisfying the following assumptions.
\begin{assumption} \label{a_acyclic_game}
The game $\Gamma$ is weakly acyclic. That is, for any $a\in\calA$, there exists a best-response path that converges to a pure-strategy Nash equilibrium.
\end{assumption}
A discussion of weakly acyclic games can be found in \cite{Young_2004}.
\begin{assumption} \label{distinct_actions_assumption}
All pure-strategy Nash equilibria of the game $\Gamma$ are strict.
\end{assumption}
We remark that Assumption \ref{distinct_actions_assumption} is generic in the sense that if the number of players and actions are fixed, then the set
of utility functions for which Assumption \ref{distinct_actions_assumption} fails to hold is a closed set of Lebesgue measure zero within the space of all possible
utility functions \cite{harsanyi1973oddness}.
\begin{assumption}\label{a_measurability}
Let $\{\mathcal{F}_{t}\}_{t\geq 1}$ be a filtration (sequence of increasing $\sigma$-algebras) with $\mathcal{F}_{t} := \sigma(\{a_s\}_{s=1}^t)$.
The strategy estimate $\hat \sigma_{j,t}^{i}\in\triangle(\calA_j)$ that agent $i$ has of the strategy $\sigma_{j,t}$ of agent $j$ is measurable with respect to $\mathcal{F}_t$.
\end{assumption}

Assumption \ref{a_measurability} means that the strategy estimates of agent $i$ are restricted to be a function of the history of play.




\subsection{Inertial Best Response Dynamics: Convergence Under Informational Limitations} \label{sec_general_result}
The following condition provides a basic sufficient condition, under which convergence to pure NE may still be ensured in distributed settings when players' ability to gather information is restricted by some sparse interagent communication graph.
\begin{condition}\label{assumption_prediction_convergence}
There exist a positive integer $T\in\N_+$ such that if any action $a\in\calA$ is repeated consecutively for $\tilde T\geq T$ stages (i.e., $a_s = a$ for $s=t,\ldots,t+\tilde T-1$), then $\arg\max_{\alpha_i\in \calA_i} \bru_{i,t+\tilde T-1}(\alpha_i) = \arg\max_{\alpha_i\in\calA_i} u(\alpha_i,a_{-i})$ for all $i\in\calN$.
\end{condition}

The condition above means that if players repeat an action for a sufficient number of stages, then players are able to learn to best respond to the actions played by others.
In the context of distributed algorithms, this relatively mild condition will effectively ensure that information is tracked sufficiently well so that the best response learning process can lock into a pure NE strategy when one is played.

The following theorem establishes that inertial best response dynamics converge under condition \ref{assumption_prediction_convergence}.
\begin{theorem} \label{thrm_main_result}
Let $\{a_t\}_{t\geq 1}$ be a sequence of actions generated by inertial best response dynamics. Suppose Assumptions \ref{a_acyclic_game}--\ref{a_measurability} and Condition \ref{assumption_prediction_convergence} hold. Then the action sequence $\{a_t\}_{t\geq 1}$ converges to a pure-strategy NE of the game $\Gamma$, almost surely. Moreover, let $\tau\in[1,\infty]$ be a random variable indicating the round number in which the action sequence $a_t$ is absorbed to a pure-strategy NE. Then $E(\tau)<\infty$.
\end{theorem}

We note that Theorem \ref{thrm_main_result} may be seen as a robust version of Young's result for finite memory better reply processes (\cite{Young_2004}, Theorem 6.2), that extends to infinite memory processes.
Better reply processes are a generalization of best reply processes in which players choose actions with utility better than the past average \cite{Young_2004}. In order to simplify the presentation, in this note we only consider the simpler case of best reply dynamics.

We will prove Theorem  \ref{thrm_main_result} using a similar approach to \cite{Young_2004}.
Lemma \ref{lemma_NE_absorption} shows that pure-strategy Nash equilibria are absorbing, and Lemma \ref{lemma_positive_absorb_probability} shows that the probability of reaching such an absorbing state is uniformly bounded from below. Together these prove Theorem \ref{thrm_main_result}.

%
\begin{lemma}[absorption property] \label{lemma_NE_absorption}
Let $\{a_t\}_{t\geq 1}$ be a sequence of actions generated by an inertial best response algorithm. Suppose Assumptions \ref{a_acyclic_game}--\ref{a_measurability} and Condition \ref{assumption_prediction_convergence} hold.
There exists a $T_1 \in \mathbb{N}_+$ such that if $a^* \in \calA$ is any pure-strategy Nash equilibrium, and if $a^*$ is played in $T_1$ consecutive stages, i.e., $a_s = a^*, ~\forall s = t,\ldots,t+T_1-1$, then
$a_{t + \tau} = a^*$ for all $\tau \geq 0$.
\end{lemma}
\begin{myproof}
Let $T$ be as in Condition \ref{assumption_prediction_convergence}, and let $T_1 \geq T$. Suppose $a^*$ is a pure Nash equilibrium and $a_s = a^*$ for $s = t,\ldots, t+T_1-1$. Then by Condition \ref{assumption_prediction_convergence},
$\argmax_{a'_i \in \calA_i} \hat u_{i,t+\tilde T}(a'_i) = \argmax_{a'_i \in \calA_i} u_i(a'_i,a^*_{-i}), ~\forall i$.
Moreover, by Assumption \ref{distinct_actions_assumption}, the set $\argmax_{a'_i \in \calA_i} u_i(a'_i,a^*_{-i}) = \{a_i^*\}$ is a singleton for each $i$.
Thus, the action $a^*$ is repeated in stage $t + T_1$. Inductively, we see that $a_{t+\tau} = a^*$ for all $\tau \geq 0$.
\end{myproof}


\begin{lemma}[positive probability of absorption]\label{lemma_positive_absorb_probability}
Let $\{a_t\}_{t\geq 1}$ be a sequence of actions generated by an inertial best response algorithm. Suppose Assumptions \ref{a_acyclic_game}--\ref{a_measurability} and Condition \ref{assumption_prediction_convergence} hold.
Let $T_1$ be as in Lemma \ref{lemma_NE_absorption}, and let $T_2 \geq T_1$ be given. Let $t$ be the current stage of the repeated play. Define the event
\begin{align*}
E_t:= \{a_\tau = a^* \mbox{ for some pure strategy NE } a^*\\
 \mbox{ for all } \tau \in\{ t',t'+1,\ldots,t'+T_2-1\},\\
 \mbox{ for some } t' \in \{t,\ldots,t+T_2|\calA|\} \}.
\end{align*}
There exists an $\epsilon = \epsilon(T_2)  > 0$ such that $\mathbb{P}(E_t\vert \mathcal{F}_t) > \epsilon$ for all $t\geq 1$.
\end{lemma}
\begin{myproof}
The proof follows along the lines of the proof of Theorem 3.1 in \cite{Marden_et_al_2009}.
By Condition \ref{assumption_prediction_convergence}, for any $t'\geq 1$, if any action $a\in \calA$ is repeated consecutively from stage $t'$ to stage $t'+T_2-1$, then
$\argmax_{\alpha_i \in\calA_i} \hat u_{i,t'+T_2-1}(\alpha_i) = \argmax_{\alpha_i \in\calA_i} u_i(\alpha_i,a_{-i})$.
Let $a^0 = a_t$. Conditioned on $\mathcal{F}_{t}$, the action $a^0$ will be played repeatedly in $T_2$ consecutive stages with probability at least $\epsilon_1 := \rho^{n(T_2-1)}>0$. Supposing this occurs, then at stage $\tau = t+T_2-1$,
$\argmax_{\alpha_i \in\calA_i} \hat u_{i,\tau}(\alpha_i) = \argmax_{\alpha_i \in\calA_i} u_i(\alpha_i,a^0_{-i})$.
At this point, either no players can improve their utility (in which case we are at a pure NE),
or at least one player can improve their utility. If the latter is the case then, conditioned on $\mathcal{F}_{t+T_2-1}$, with probability at least $\epsilon_2 := \rho^{n-1}(1-\rho)$, exactly one player $i$ chooses to take a best response and improves their utility, and all others continue to play $a^0_{-i}$. Call the new action profile $a^1$. Continuing in this manner, we can construct a sequence of actions $a^0,a^1,\ldots,a^m$ (terminating with at most $m = |\calA|$) such that $a^m$ is a pure-strategy Nash equilibrium. Conditioned on $\mathcal{F}_{t}$, the probability of this action sequence occurring (and then the final action $a^m$ being played for $T_2$ consecutive stages) is bounded from below by $\epsilon := (\epsilon_1\epsilon_2)^{|\calA|}\rho^{T_2-1}$.
\end{myproof}

We now prove Theorem \ref{thrm_main_result}.

\begin{myproof}
Let $T_2$ be as in Lemma \ref{lemma_positive_absorb_probability}.
By Lemma \ref{lemma_NE_absorption}, if a pure NE action $a^*$ is played in $T_2$ consecutive stages, then $a^*$ will be played in all consecutive stages. By Lemma \ref{lemma_positive_absorb_probability}, the probability of reaching such an ``absorbing state'' is uniformly lower bounded by some $\epsilon >0$. Thus, the process is absorbed to a pure NE almost surely in finite time, and $E(\tau)<\infty$ (\cite{williams1991probability}, p.233).
\end{myproof}


%
\section{Distributed Fictitious Play with Inertia}\label{sec_D-FP}
In this section we will study a variant of the classical FP algorithm in which the best response of classical FP is augmented with an inertia term, and inter-agent communication is restricted to a graph.
We begin by reviewing the centralized FP with fading memory and inertia. We will develop a distributed variant of this algorithm that operates in network-based settings satisfying Assumption \ref{a_comm_graph}.

%
\subsection{Fictitious Play with Inertia and Fading Memory} \label{sec_FP_FM_and_inertia}
A review of the classical FP algorithm can be found in  \cite{Fudenberg_Levine_1998,Young_2004}.
The FP with inertia algorithm is defined as follows.
Given an action $a_i\in \calA_i$, let $\Psi(a)\in \triangle(\calA_i)$ be the degenerate probability distribution placing mass 1 on the action $a_i$.
Let $f_{i,t}\in \calR^{|\calA_i|}$ denote the weighted empirical distribution (or just \emph{empirical distribution}) of player $i$.
Formally, $f_{i,t}$ may be defined recursively by letting $f_{i,1} = \Psi(a_{i,1})$ and for $t\geq 1$ letting
\begin{equation} \label{empirical_distribution_recursion_fading}
f_{i,t+1} = (1-\alpha)f_{i,t} + \alpha \Psi(a_{i,t+1}),
\end{equation}
where $\alpha \in (0,1]$ is a step-size parameter.

Let the joint weighted empirical distribution profile (or joint empirical distribution) be given by $f_{t} := (f_{1,t},\ldots,f_{n,t})$. The weighted empirical distribution is said to have ``fading memory'' because it places greater weight on recent events.\footnote{This is a consequence of the fact that $\alpha$ is a time-invariant constant. In classical FP, the associated constant is permitted to be time-varying with $\alpha_t = \frac{1}{t+1}$, which results in $f_{i,t}$ being a histogram placing equal weight on the events from all previous rounds. While the use of inertia is essential to the structure of our proofs, the use of fading memory is less critical. It is possible that the results still hold using a time-varying step size $\alpha_t$, (e.g., \cite{Marden_et_al_2009}, Section II-E); however, the assumption of fading memory simplifies the analysis.}

In fictitious play with fading memory and inertia, each player chooses their next-stage action according to the rule
\begin{equation}
a_{i,t+1} \in
\begin{cases}
\argmax_{\alpha_i \in \calA_i} u_i(\alpha_i,f_{-i,t}) & \mbox{ with prob. } 1-\rho,\\
a_{i,t} & \mbox{ with prob. } \rho,
\end{cases}
\end{equation}
where $\rho\in (0,1)$ is some predefined ``inertial constant'' and the probability is conditioned on $\mathcal{F}_{t-1}$ (see Assumption \ref{a_measurability}).
The constant $\rho\in (0,1)$ adds a form of ``inertia'' by increasing the probability that the current action will be repeated in upcoming stages.

FP with inertia can be shown to converge to pure NE in weakly acyclic games satisfying Assumption \ref{distinct_actions_assumption}, which includes almost all potential games. Examples of interest include any multi-player engineered system with a global objective, e.g., power control in communication networks \cite{candogan2010near,hicks2004game}, sensor coverage \cite{martinez2007motion}, and wind energy harvesting \cite{marden2012surveying}.

In the distributed setting, players may lack sufficient information to precisely compute the empirical distribution $f_{i,t}$. Let $\hat f^i_{j,t}$ be an estimate that player $i$ maintains of $f_{j,t}$. Let $\hat f^i_t = (\hat f^i_{1,t},\ldots,\hat f^i_{n,t})$ be an estimate that player $i$ maintains of the empirical distribution profile $f_t$.


\subsection{Distributed FP Algorithm}
For each $j\in\calN$, let  $W_j=(w_{j,k}^i)_{i,k}\in\R^{n\times n}$ be a weight matrix to be used by player $j\in\calN$ in the distributed algorithm. The distributed FP with inertia algorithm is given below. We assume that players are in a distributed setting such that Assumption \ref{a_comm_graph} holds. Thus, the only information available to players is observations of their own actions, and whatever information is transmitted to them by their neighbors in previous rounds.

\begin{algorithm}\label{algo_dfp_example} \normalfont
$~$\\
\noindent \textit{Initialize}\\
(i) Let $\rho \in (0,1)$ be fixed.
For each $i$, let the initial action $a_{i,1}$ be chosen arbitrarily, let $f_{i,1} = \Psi(a_{i,1})$, and let $\hat f_{i,1}^i = f_{i,1}$.
For $j\not = i$, let $\hat f_{j,1}^i = w_{j,j}^i f_{j,1}$ if $j\in \mathcal{N}_i$ and
$\hat f_{j,1}^i = 0$ otherwise, where $w_{j,j}^i$ is a weight constant (see step (iv) and Lemma \ref{cor_dist_error} below). \\


\noindent \textit{Iterate} ($t\geq 1$)\\
(ii) Each agent $i$ chooses their next-stage action according the rule
\begin{equation}
a_{i,t+1} =
\begin{cases}
\argmax_{\alpha_i \in \calA_i} u_i(\alpha_i,\hat f^i_{-i,t}) & \mbox{ with prob. } 1-\rho,\\
a_{i,t} & \mbox{ with prob. } \rho.
\end{cases}
\end{equation}\\

\noindent (iii) Each player $i$ updates their personal empirical distribution $f_{i,t}$ according to \eqref{empirical_distribution_recursion_fading}.\\

\noindent (iv) For each $i$, $\hat f^i_{j,t}$ is updated as
\begin{align} \label{empirical_distribution_estimate_recursion}
\hat f^i_{j,t+1} =   \sum_{k\in \calN_i}   w^i_{j,k}\left( \hat f^k_{j,t} + \left(f_{j,t+1} - f_{j,t} \right)\chi_{\{k=j\}} \right),
\end{align}
where $\chi_{\{k=j\}}$ is the characteristic function defined by $\chi_{\{k=j\}} = 1$ if $k=j$ and $\chi_{\{k=j\}} = 0$ otherwise, and where $w^i_{j,k}$ is the weight that player $i$ attributes to $k$'s estimate of $j$'s empirical frequency (see Lemma \ref{cor_dist_error}.)
\end{algorithm}

\subsection{Distributed FP with Inertia: Convergence Analysis} \label{sec_D-FP_analysis}
The following result establishes the convergence of Algorithm \ref{algo_dfp_example}.
\begin{theorem} \label{thrm_dfp_specific}
Suppose Assumptions \ref{a_comm_graph}, and \ref{a_acyclic_game}--\ref{a_measurability} hold.
Let $W_j \in R^{n\times n}$, $j\in\mathcal{N}$ be a \emph{weight matrix} with the $i,k$-th entry given by $W_j(i,k) = w^i_{j,k}$.
Assume that the matrix $W_j$ is row stochastic with sparsity conforming to the communication network $G$. Assume the $j$-th diagonal entry satisfies $w^j_{j,j}=1$ for each $j\in\mathcal{N}$. Let $P_j$ be the matrix obtained by removing the $j$-th row and column from $W_j$. Assume $P_j$ is irreducible and substochastic in the sense that at least one row sum of $P_j$ is strictly less than $1$.
Then Algorithm \ref{algo_dfp_example} converges to a pure NE, almost surely.
\end{theorem}

We remark that conditions on the weight matrix $W_j$ above are closely related to those found in the literature on higher-dimensional consensus \cite{khan2010higher}.

Note that Algorithm \ref{algo_dfp_example} is an inertial best response process, hence it converges to pure NE a.s. if Condition \ref{assumption_prediction_convergence} holds. By Lipschitz continuity of $u_i$, if
$\|f_{i,t} - f_{i,t}^j\|\to 0$ for all $i,j$ then $|u_i(\alpha_i,f_{-i,t}) - u_i(\alpha_i,\hat f_{-i,t}^i)|\to 0$. Thus, the following Lemma shows that under the hypotheses of Theorem \ref{thrm_dfp_specific}, Condition \ref{assumption_prediction_convergence} is satisfied. Theorem \ref{thrm_dfp_specific} then follows from Theorem \ref{thrm_main_result} and Lemma \ref{cor_dist_error}.

\begin{lemma} \label{cor_dist_error}
Assume the hypotheses of Theorem \ref{thrm_dfp_specific} hold.
Suppose $\{a_t\}_{t\geq 1}$ is generated according to Algorithm \ref{algo_dfp_example}.
Then, for any $\epsilon >0$ there exists $T\in\mathbb{N}_{+}$ such that if players repeat any action $a^*\in \calA$ for $\tilde T\geq T$ consecutive stages (i.e., $a_s = a^*,~s=t,\ldots,t+\tilde T-1$) then $\|\hat f^i_{t+\tilde T-1} -  f_{t+\tilde T-1}\| < \epsilon$.
\end{lemma}

The proof of Lemma \ref{cor_dist_error} is given in the appendix.
\begin{remark}
We note that the techniques used to prove convergence of Algorithm \ref{algo_dfp_example} to pure NE are flexible and are not restricted to the information dissemination scheme used in step (iv) of the algorithm.
In particular, any information dissemination scheme can be used in step (iv) so long as a corresponding result analogous to Lemma \ref{cor_dist_error} holds.
\end{remark} 


%
\section{Distributed JSFP with Inertia}\label{sec_D-JSFP}
FP can be difficult to implement in practice due to the high computational and memory requirements. Joint Strategy FP (JSFP) with inertia, introduced in \cite{Marden_et_al_2009}, is a variant of FP developed for large-scale games that has relatively low computational complexity and low information overhead requirements.
In this section we study a distributed variant of JSFP with inertia
(referred to hereafter as distributed JSFP)
for use in networked settings satisfying Assumption \ref{a_comm_graph}.\footnote{A related variant of JSFP---termed Average Strategy FP (ASFP)---is studied in \cite{xiao2013average}. However, ASFP differs fundamentally from distributed JSFP in that (i) ASFP assumes instantaneous and perfect information dissemination by an oracle, and (ii) distributed JSFP uses a projection operation to make sense of the notion of players ``assuming'' that the average congestion profile represents choices taken by agents.}

The variant of JSFP that we study is applicable within the class of congestion games---a subset of the more general class of weakly-acyclic games (see Assumption \ref{a_acyclic_game}). This restriction comes as a consequence of the manner in which information is aggregated over the communication network.
Thus, while distributed JSFP operates with lower complexity and communication overhead than distributed FP (Section \ref{sec_D-FP}), distributed JSFP is applicable within a narrower class of games than distributed FP.

The class of congestion games is introduced in Section \ref{sec_JSFP_congestion_games}, the distributed JSFP algorithm is presented in Sections \ref{sec_D-JSFP_setup}--\ref{sec_D-JSFP_algo}, and convergence of the algorithm is analyzed in Section \ref{sec_D-JSFP_analysis}.

\subsection{Congestion Games} \label{sec_JSFP_congestion_games}
Let $R=\{1,\ldots,m\}$ denote a set of resources. For each $i\in \mathcal{N}$, let $\calA_i \subseteq 2^{R}$, where $2^{R}$ denotes the power set of $R$. In particular, an action choice $a_i$ indicates a subset of resources being utilized by player $i$.

In a congestion game, the cost associated with using a resource is dependent on the total number of players using the same resource. For each $r\in R$, $a\in \calA$, let $N_r(a)\in \mathbb{N}$ denote the number of players using resource $r$ under the action profile $a$. More generally, for a subset of players $\calK \subseteq \mathcal{N}$, the number of players in $\calK$ utilizing resource $r$ given $(a_j)_{j\in\calK}$, is given by
\begin{equation}
N_r((a_j)_{j\in\calK}) := \sum_{j\in \calK} \ones(r\in a_j ).
\end{equation}
where $\ones(r\in a_j) = 1$ if $r\in a_j$ and $\ones(r\in a_j) = 0$ otherwise.
Given a subset of players $\calK$, and a corresponding set of actions $(a_j)_{j\in\calK}$, we represent the number of players using each resource by $N((a_j)_{j\in\calK})$, where $N:\prod_{j\in\calK} A_j\to \mathbb{N}^m$ is a mapping with the $r$-th entry in $N((a_j)_{j\in \calK})$ given by $N_r((a_j)_{j\in \calK})$.

For $r\in R$ and $k\in\mathbb{N}$, let $c_r(k)$ be the cost associated with using resource $r$, when there are precisely $k$ players simultaneously using the resource. For $a_i\in\calA_i$ and $N_r(a_{-i})\in \mathbb{N}$, let the utility of player $i$ be given by
\begin{align}
u_i(a_i,a_{-i}) & = -\sum_{r\in a_i}c_{r}(N_r(a))\\
& = -\sum_{r\in a_i}c_{r}(N_r(a_{-i})+N_{r}(a_i))
\end{align}
where we have written $N_r(a) = N_r(a_{-i})+N_{r}(a_i)$ explicitly to emphasize dependence of the utility on ``self action'' $a_i$ and actions of other players $a_{-i}$. Note that within the class of congestion games, players do not need to precisely know the full action profile $a = (a_1,\ldots,a_n)\in \calA$ to compute their utility. It is sufficient for each player to have knowledge of $N(a_{-i})\in\mathbb{N}^m$ and their own action $a_i \in \calA_i$. In this context, we sometimes express the utility function using the abuse of notation $u_i(a_i,N(a_{-i})) =  u_i(a_i,a_{-i})$.

In the following, we use this property of the utility functions in congestion games to design the distributed JSFP algorithm which has a lower communication overhead than distributed FP.

\subsection{Distributed JSFP Setup} \label{sec_D-JSFP_setup}
Assume players repeatedly face off in a congestion game.
We define $\zeta_{i,t}(r)$ to be a (fading-memory) weighted average used to track the amount of congestion induced on resource $r$ by the actions of (only) player $i$. In particular, let $\zeta_{i,t}(r)$ be defined recursively by $\zeta_{i,1}(r):= N_r(a_{i,1})$, and for $t\geq 1$,
\begin{equation} \label{eqn_zeta_update}
\zeta_{i,t+1}(r) := (1-\alpha)\zeta_{i,t}(r) + \alpha N_r(a_{i,t}),
\end{equation}
where $\alpha \in (0,1]$ is a weight parameter inducing a fading-memory effect (cf. \eqref{empirical_distribution_recursion_fading} and subsequent discussion).

Furthermore, define $\zeta_{i,t}\in \mathbb{R}^m$ to be the vector stacking $(\zeta_{i,t}(r))_{r\in R}$---this is a vectorized representation of the congestion induced by player $i$ on any given resource.

Define $\zeta_{t}(r):=\sum_{j\in\calN} \zeta_{j,t}(r)$---this represents the congestion induced on resource $r$ by the actions of \emph{all} players. Note this can also be expressed recursively as $\zeta_{t}(r) = (1-\alpha)\zeta_{t}(r) + \alpha N_r(a(t))$.

Similar to the above, let  $\zeta_{t}$ be a vector in $\mathbb{R}^m$ stacking $(\zeta_{t}(r))_{r\in R}$---this is a vectorized representation of the congestion induced by \emph{all} players on any given resource.
We refer to $\zeta_t$ as the empirical congestion distribution.

In the distributed framework, players may not have precise knowledge of $\zeta_{t}$. Instead, we assume each player $i$ maintains an estimate of $\zeta_{t}$ which we denote by $\hat \zeta^i_t \in \mathbb{R}^m$. The $r$-th term of player $i$'s estimate, $\hat \zeta^i_t(r)$, represents her estimate of the congestion at resource $r\in R$.

Finally, in order to rigorously define distributed JSFP, we require the following notion of a projection. For a vector $v\in\mathbb{R}^m$ define $P(v)$ to be a projection of $v$ onto the set of non-negative $m$-dimensional integer-valued vectors $\mathbb{N}^m$; formally, for $1\leq r\leq m$, let $P(v,r) := z$ for the unique $z\in \mathbb{N}$ satisfying $z-\frac{1}{2} \leq v(r) < z+\frac{1}{2}$. Let $P(v)$ be the vector stacking $\{P(v,r)\}_{r\in R}$.

\subsection{Distributed JSFP Algorithm} \label{sec_D-JSFP_algo}
Let $W \in R^{n\times n}$ be a \emph{weight matrix} to be used in the distributed algorithm with the $i,k$th entry given by $w^i_{k}$.  We assume that players are in a distributed setting such that Assumption \ref{a_comm_graph} holds.
The distributed JSFP algorithm is given as follows.

\begin{algorithm}\label{algo_jsfp_example} \normalfont
$~$\\
\noindent \textit{initialize}\\
(i) Let $a_{i,1}$ be arbitrary for all $i$. Let $\hat \zeta^i_1 = N(a_{i,1})$ for all $i$.

\noindent \textit{iterate} $(t\geq 1)$\\
(ii) Let $\hat \zeta^i_{-i,t} = \hat \zeta^i_{t} - \zeta_{i,t}$. For each player $i$, the next-stage action is chosen according to the rule
$$a_{i,t+1} \in
\begin{cases}
\argmax_{\alpha_i \in \calA_i} u_i(\alpha_i,P(\hat \zeta_{-i,t}^i)), & \mbox{ w.p. } 1-\rho\\
a_{i,t}, & \mbox{ w.p. } \rho.
\end{cases}
$$

\noindent(iii) Update $\zeta_{i,t+1}$ according to \eqref{eqn_zeta_update}.

\noindent(iv) Each player $i$ updates their estimate of $\zeta_t$ as:
\begin{equation} \label{JSFP_distributed_dynamics}
\hat \zeta_{t+1}^i := \sum_{k\in\calN_i} w^i_{k} \left( \hat \zeta_{t}^k + \zeta_{k,t+1} - \zeta_{k,t} \right)
\end{equation}
where $w^i_k$ denotes the weight that player $i$ places on the information received from player $k$ (see Lemma \ref{lemma_JSFP1}).
\end{algorithm}

We remark that in distributed JSFP players only share a vector with $m$ integer values with their neighbors where $m$ is the cardinality of the set of resources $R$. In comparison, the distributed FP algorithm high higher memory and communication overhead requirements in that players share their estimate of each agent's empirical frequency implying that they share $n\times m$ values at each step.
Furthermore, in classical JSFP \cite{Marden_et_al_2009} it is assumed that information is provided to players by an oracle, whereas in the distributed variant above, the algorithm explicitly handles information dissemination.

\subsection{Distributed JSFP: Convergence Analysis} \label{sec_D-JSFP_analysis}
The following theorem gives the convergence result for distributed JSFP with inertia.
\begin{theorem} \label{thrm_JSFP_specific}
Assume Assumptions \ref{a_comm_graph}, \ref{distinct_actions_assumption}--\ref{a_measurability} hold and that the matrix $W$ is doubly stochastic, aperiodic, and irreducible. Then the distributed JSFP process converges to a pure-strategy NE, almost surely.
\end{theorem}

In order to prove Theorem \ref{thrm_JSFP_specific}, we begin by showing the following lemma.
\begin{lemma} \label{lemma_JSFP1}
Let $\{a_t\}_{t\geq 1}$ and $\{\hat \zeta^1_t,\ldots,\hat \zeta^n_t\}_{t\geq 1}$ be generated according to a distributed JSFP process, and let $\{\zeta_t\}_{t\geq 1}$ be as defined in \eqref{eqn_zeta_update}.
Assume the hypotheses of Theorem \ref{thrm_JSFP_specific} hold.
There exists a $\tilde T\geq 1$ such that if any action $a^*$ is repeated in $  \tilde T\geq T$ consecutive stages then $|\hat \zeta^i_{t+ \tilde T}(r) - N_r(a^*)| <\frac{1}{4}$ for every $r\in R$.
\end{lemma}

The proof of Lemma \ref{lemma_JSFP1} is similar to the proof of Lemma \ref{cor_dist_error} (see appendix) and is omitted here for brevity.

Given Lemma \ref{lemma_JSFP1}, the following lemma shows that if any action $a^*$ is repeated in sufficiently many stages, then each player's estimate $\hat \zeta^i_{-i,t}$ may be brought sufficiently close to the congestion profile $N(a_{-i}^*)$ to ensure convergence of the process.

\begin{lemma}\label{lemma_JSFP2}
Let $\{a_t\}_{t\geq 1}$ and $\{\hat \zeta^1_t,\ldots,\hat \zeta^n_t\}_{t\geq 1}$ be generated according to a distributed JSFP process and let $\{\zeta_{i,t}\}_{i\in\calN,t\geq 1}$ be as defined in \eqref{eqn_zeta_update}.
Assume the hypotheses of Theorem \ref{thrm_JSFP_specific} hold. There exists a $T\geq 1$ such that if any action $a^*$ is repeated in $\tilde T \geq T$ consecutive stages then $|\hat \zeta^i_{-i,t+\tilde T}(r) - N_r(a^*_{-i})| <\frac{1}{2}$ for every $r\in R$.
\end{lemma}
\begin{myproof}
Let $\tilde T \geq 0$ and note that
\begin{align}
& |\hat \zeta^i_{-i,t+\tilde T}(r) - N_r(a^*_{-i})| \\
& = |(\hat \zeta^i_{t+\tilde T}(r) - \zeta_{i,t+\tilde T}(r)) - (N_r(a^*)-N_r(a^*_{i}))|\\
& \leq |\hat \zeta^i_{t+\tilde T}(r) - N_r(a^*)| + |\zeta_{i,t+\tilde T}(r) - N_r(a^*_{i})|.\label{eqn_JSFP2_eq1}
\end{align}
By Lemma \ref{lemma_JSFP1}, we may choose $T'$ such that if $a^*$ is repeated in $\tilde T \geq T'$ consecutive stages, there holds $|\hat \zeta^i_{t+\tilde T}(r) - N_r(a^*)| < \frac{1}{4}$. Note also that $|\zeta_{i,t+\tilde T}(r) - N_r(a^*_{i})| \rightarrow 0$ as $\tilde T \rightarrow \infty$ (this follows from \eqref{eqn_zeta_update}), and thus we may choose $T''$ such that for $\tilde T \geq T''$ there holds $|\zeta_{i,t+\tilde T}(r) - N_r(a^*_{i})| < \frac{1}{4}$. Letting $T = \max\{T',T''\}$, the desired result follows from \eqref{eqn_JSFP2_eq1}.
\end{myproof}

The next lemma sets us up to prove convergence of distributed JSFP using Theorem \ref{thrm_main_result} and Condition \ref{assumption_prediction_convergence}.

\begin{lemma}\label{lemma_JSFP3}
Let $\{a_t\}_{t\geq 1}$ and $\{\hat \zeta^1_t,\ldots,\hat\zeta^n_t\}_{t\geq 1}$ be generated according to a distributed JSFP process, and let $\{\zeta_{i,t}\}_{i\in\calN,t\geq 1}$ be as defined in \eqref{eqn_zeta_update}.
Assume the hypotheses of Theorem \ref{thrm_JSFP_specific} hold.
There exists a $T\geq 1$
such that if any action $a^*$ is repeated in $\tilde T \geq T$ consecutive stages then $\argmax_{\alpha_i \in \calA_i} u_i(\alpha_i,P(\hat \zeta^i_{-i,t+\tilde T})) = \argmax_{\alpha_i \in \calA_i} u_i(\alpha_i,N(a^*_{-i}))$.
\end{lemma}
\begin{myproof}
Let $T$ be chosen as in Lemma \ref{lemma_JSFP2} so that $|\hat \zeta^i_{-i,t+\tilde T}(r) - N_r(a^*_{-i})| <\frac{1}{2}$ for every $r\in R$, $i\in\mathcal{N}$ and all $\tilde T \geq T'$. It follows that $P(\hat \zeta^i_{-i,t+\tilde T}) = N(a^*_{-i})$. Thus, $\argmax_{\alpha_i \in \calA_i} u_i(\alpha_i,P(\hat \zeta^i_{-i,t+\tilde T})) = \argmax_{\alpha_i \in \calA_i} u_i(\alpha_i,N(a^*_{-i}))$.
\end{myproof}

Finally, we note that Algorithm \ref{algo_jsfp_example} is an inertial best response process, fitting the template of Theorem \ref{thrm_main_result}, with $u_i(\alpha_i,P(\hat \zeta^i_{-i,t})) = \bru_{i,t}(\alpha_i)$ for each $i$ and each $\alpha_i \in \calA_i$.
By Lemma \ref{lemma_JSFP3}, the sequence $\{u_i(\alpha_i,P(\zeta^i_{-i,t}))\}_{t\geq 1}$ satisfies Condition \ref{assumption_prediction_convergence}. Theorem \ref{thrm_JSFP_specific} then follows from Theorem \ref{thrm_main_result}.

\begin{remark}
We note that the techniques used to prove convergence of Algorithm \ref{algo_jsfp_example} to pure NE are flexible and are not restricted to the information dissemination scheme used in step (iv) of the algorithm.
In particular, any information dissemination scheme can be used in step (iv) so long as a corresponding result analogous to Lemma \ref{cor_dist_error} holds.
We also note that while we do not consider time-varying communication networks in this note, our results can be extended to such settings so long as the estimate updates satisfy Condition \ref{assumption_prediction_convergence}.
\end{remark}



%
\section{Distributed UAV target assignment} \label{sec_simulation}

We consider the effect of ommunication network $G$ on convergence time in the example of UAV target assignment problem. We consider $n$ UAVs, and $n$ target objects. Each UAV can target one object and goal is to target all of the objects as a team. The action space is the set of objects $\{1,2\dots, n\}$ for each UAV. The payoff of UAV $i$ targeting object $k$ is inversely proportional to its distance to the object, represented by $d(i,k)$, if no other UAV is targeting object $k$,
\begin{equation} \label{utility_i_target_k}
u_i(a_i = k,a_{-i}, d(i,k)) = d(i,k)^{-1} \bbone(\sum_{j=1}^n \bbone(a_j = k) = 1)
\end{equation}
where $\bbone(\cdot)$ is the indicator function. The target assignment game with payoffs as above is a congestion game with each object representing a resource. Note that any action profile that covers all the objects is a Nash equilibrium because any unilateral deviation from such profile results in zero payoff for the deviating agent. The optimal Nash equilibrium profile minimizes the total distance while covering all the objects.

\begin{figure}\centering
\includegraphics[width=0.95\linewidth]{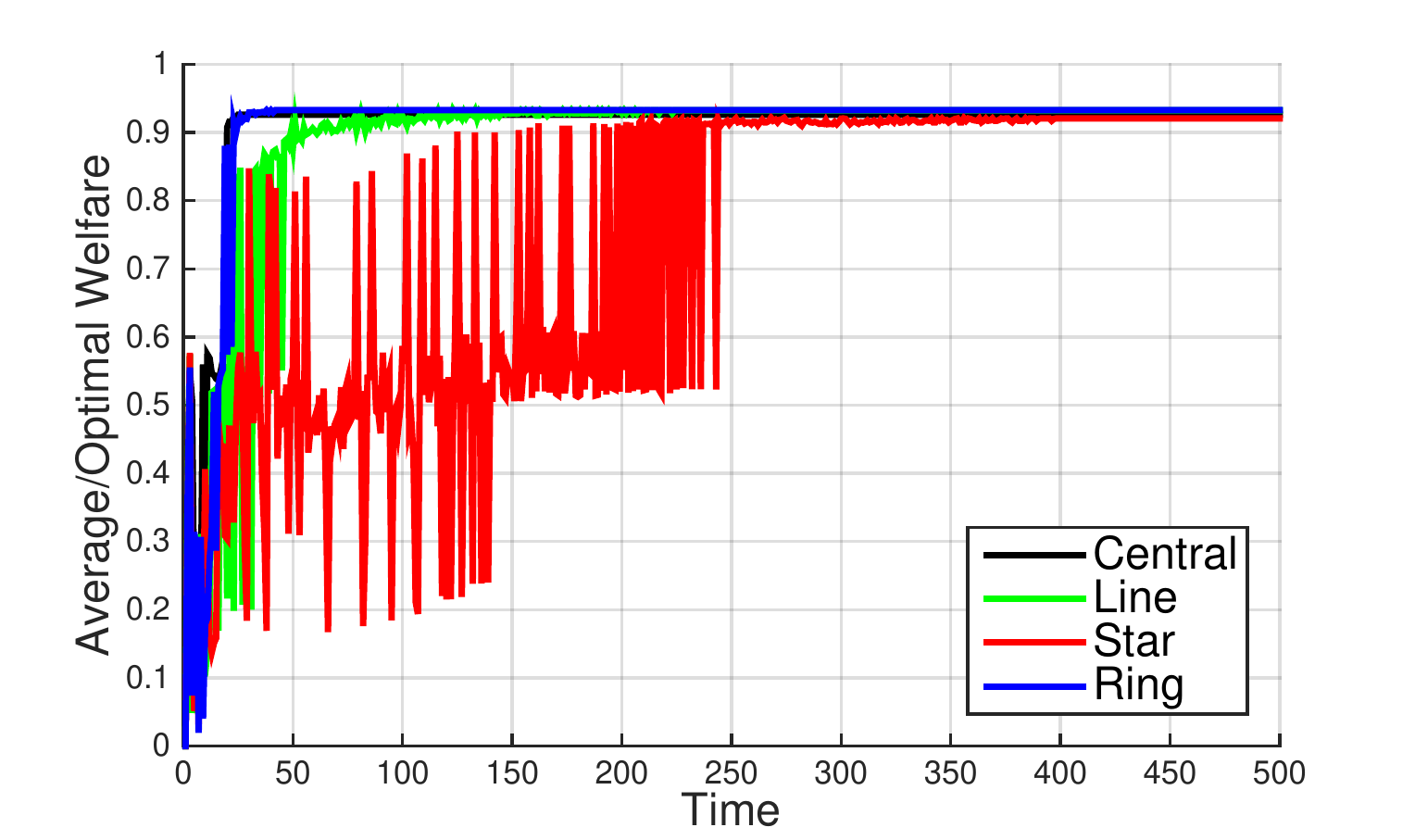}
\caption{Expected welfare normalized by optimal welfare for complete, line, star and ring networks. The expected welfare is similar for all communication networks. Convergence time to a Nash equilibrium is the fastest for the complete network and slowest for the star network.}
\label{fig_average_welfare} \end{figure}

In the numerical setup, we consider $n=5$ UAVs and $n=5$ objects with $\alpha = 0.2$ and $\delta = 0.2$. For comparison, we consider the centralized JSFP (complete network), and D-JSFP in line, ring and star communication networks. For each setting, we consider 50 runs. In Figure \ref{fig_average_welfare}, we plot sample average welfare normalized by the optimal welfare over time for each network type. Welfare at time $t$ is defined as the sum of utilities of UAVs given the action profile generated by the D-JSFP process at time $t$. The expected welfare at time $t$ is the average of welfare values at time $t$ obtained over 50 runs. Optimal welfare is the value of welfare obtained by the action profile that maximizes the sum of the utilities. Fig. \ref{fig_average_welfare} shows that the expected welfare of the Nash equilibrium reached by the algorithm is similar regardless of the communication network. However, the convergence time of the algorithm depends on the network structure where the star network has the slowest convergence time and the ring network has the fastest convergence time comparable to the centralized JSFP. 

%

We further analyze the effect of the inertia and fading constants on convergence time where we assume $\alpha \in (0.1, 0.9)$ and $\rho \in (0.1, 0.9)$. We consider increments of $0.1$ for each constant and simulate 50 runs for a given pair of $\alpha$ and $\rho$ values. Convergence time is relatively worse when $\alpha$ is high (greater than 0.7) and $\rho$ is small (less than 0.3). This worst case is when agents are sensitive to current events and often best respond. In other values of the constants, the convergence time to pure Nash equilibria are comparable for a given network structure. In general when fading constant is small ($\alpha \leq 0.3$), a broad range of the inertia constant $\rho\in(0.1,0.8)$ achieves relatively fast convergence. We show the average number of steps for convergence when $\alpha = 0.2$ with respect to different $\rho$ values in Table \ref{table:convergence_time}. 

\begin{table}[t]
\centering
\begin{tabular}{@{}l l l l l @{}}\toprule
\multicolumn{1}{ c  }{} & \multicolumn{4}{ c }{Inertia $\rho$}\\
\cmidrule{2-5}
\multicolumn{1}{ c  }{} & 0.2  & 0.4  & 0.6 & 0.8\\ \midrule
\multicolumn{1}{ c  }{Central}& 22   & 22  & 25 & 38 \\
\multicolumn{1}{ c  }{Line}& 146  & 148  & 162 & 104\\
\multicolumn{1}{ c  }{Star}& 404  & 430  & 364 & 245\\
\multicolumn{1}{ c  }{Ring}& 30  & 33  & 34 & 37\\
\bottomrule
\vspace{0pt}
\end{tabular}
%
%
%
%
%
\caption{D-JSFP Algorithm: average convergence time }
\label{table:convergence_time}
\end{table}

%
\section{Conclusion} \label{conclusion}
We have studied general inertial best response dynamics for learning pure-strategy NE in distributed multi-agent systems.
Subsequently, we have studied two important cases of inertial best response dynamics (FP and JSFP); we have derived distributed variants of both algorithms and derived sufficient conditions for convergence. Results were corroborated with
a simulation example of an $n$-UAV target assignment problem.
In future research, it may be interesting to investigate the extension of these techniques to develop distributed implementations of related algorithms such as no-regret algorithms, e.g., \cite{marden2007regret}.

{\small
%
\section*{Appendix}
\label{section_distributed_averaging}
Consider a network of $n$ nodes connected through a communication graph $G=(V,E)$.  The graph is assumed to be strongly connected.
For $t=1,2,\ldots$ let $x_1(t)\in\mathbb{R}$ denote a value held by node $1$ at time $t$. The objective is for all nodes to track as closely as possible the value $x_1(t)$.\footnote{In general, the objective may be to track the value $x_j(t)$ held by an arbitrary node $j$. Here, we only consider tracking $x_1(t)$, however, the general case is recovered by a permutation of the node labels.}
Let $\epsilon(t):= |x_1(t+1) - x_1(t)|$ and assume that:
\begin{assumption} \label{assumption_err_bound}
There exists a $B>0$ such that $\epsilon(t) < B$ for all $t\in\mathbb{N}$.
\end{assumption}

Let $\hat x_i(t)$ be the estimate player $i$ maintains of $x_1(t)$. 
We make the following assumption pertaining to the initial error in players' estimates:
\begin{assumption}
$\hat x_i(0) - x_1(0) = 0 ~\forall i$.
\label{assumption_initial_error}
\end{assumption}

Let $\hat x(t) = (\hat x_1(t),\ldots, \hat x_n(t))\in \mathbb{R}^n$ be a vector stacking all players' estimates, with $\hat x_1(t) = x_1(t)$ (i.e., node $j$ knows its own value).
Suppose the estimates are updated according to the following recursion:
\begin{equation}
\hat x(t+1) = W\left( \hat x(t) + e_1(x_1(t+1) - x_1(t)) \right),
\label{update_rule1}
\end{equation}
where $e_1\in\mathbb{R}^n$ is the 1-st canonical vector, and where the matrix $W\in\mathbb{R}^{n\times n}$ satisfies
\begin{assumption} \label{assumption_W_matrix}
$W$ is row stochastic with sparsity conforming to $G$. Furthermore, $W$ may be decomposed as
$$
W =
\begin{pmatrix}
1 ~ 0\\
b ~ P
\end{pmatrix}
$$
where $b\in \mathbb{R}^{n-1}$, $b\not= 0$ and $P$ is irreducible (cf. \cite{khan2010higher}).
\end{assumption}
Note that in since $W$ is row stochastic, $b\not=0$ if and only if $P$ is substochastic in the sense that at least one row sum of $P$ is less than 1.

The following lemma gives a bound on the error in the agents' estimates of $x_1(t)$.
\begin{lemma} \label{lemma_distributed_averaging}
Suppose Assumptions \ref{assumption_err_bound}--\ref{assumption_initial_error} hold, and let the sequence $\{\hat x(t)\}_{t=1}^\infty$ be computed according to \eqref{update_rule1}. Suppose there exists a $t^* \geq 1$ and $T\geq 1$ such that
$\sum_{t=t^*}^{t^*+T-2} \epsilon(t) \leq 1$ and $\{\epsilon(t)\}_{t=t^*}^{t^*+T-2}$ is decreasing. Then the error at time $t^* + T-1$ is bounded as,
$$\| \hat x(t^* + T-1) -  x_1(t^* + T-1) \ones \| \leq \frac{n+1}{1-\lambda}\left(\frac{1}{T} + B\lambda^{T}\right), $$
where $\lambda := \sup\limits_{\|y\| = 1} \| Py \|<1$.
\end{lemma}
\begin{myproof}
Let $y(t) := \hat x(t)- x_1(t) \ones$. Let $\delta(t) := [x_1(t+1) - x_1(t)]e_1 - [x_1(t+1) - x_1(t)]\ones$. Subtracting $x_1(t+1)\ones$ from both sides of \eqref{update_rule1} we get
\begin{align}
y(t+1) & = W\left(\hat x(t) + [x_1(t+1) - x_1(t)]e_1\right) - x_1(t+1)\ones\\
& = W\left(\hat x(t) + [x_1(t+1) - x_1(t)]e_1 - x_1(t+1)\ones\right)
\end{align}
where, in the second line, we may bring $x_1(t+1)\ones$ inside the matrix multiplication due to the row stochasticity of $W$. Now we add and subtract $x_1(t)$ and use the definitions of $y(t)$ and $\delta(t)$ to get
$
y(t+1) = W\left( y(t) + \delta(t)  \right).
$
Inductively, this gives $y(t+1) = \sum_{s=0}^t W^{s+1}\delta(t-s) + W^{t+1}y(0)$. By Assumption  \ref{assumption_initial_error} we have $y(0) = 0$, and hence $y(t+1) = \sum_{s=0}^t W^{s+1}\delta(t-s)$.
By the triangle inequality we have
\begin{align}\label{lemma_DA_eq1}
\|y(t+1)\| \leq \sum_{s=0}^t \|W^{s+1}\delta(t-s)\|.
\end{align}
\vskip-5pt
Again using the triangle inequality, we establish a bound on $\|\delta(t)\|$:
\begin{align}
\|\delta(t)\| & \leq \|[x_1(t+1) - x_1(t)]e_1\| + \|[x_1(t+1) - x_1(t)]\ones\|\\
& \leq \epsilon(t) + n\epsilon(t) = (n+1)\epsilon(t).
\label{lemma5_delta_ineq}
\end{align}
Let $\overline{W} := W - \ones e_1^T$. In block form we have $\overline{W} = [0 ~ \ldots ~ 0;~ (b-\ones) ~~ P]$.
Due to the special block form of $\overline{W}$, the spectrum $\sigma(\overline{W})$\footnote{In Section \ref{sec_prelims} the symbol $\sigma$ was used to represent a mixed strategy. In keeping with standard conventions, we use $\sigma$ here to denote the spectrum of a matrix, where the distinction is clear from the context.} of $\overline{W}$ consists precisely of $\{\sigma(P) \cup \{0\}\}$. Hence, the spectral radius of $\overline{W}$ coincides with that of $P$. In particular, the spectral radius of $\overline{W}$ is given by $\lambda$.
Since $P$ is substochastic, we have $\lambda <1$.


Substituting $W = \overline{W} + \ones e_1^T$ in \eqref{lemma_DA_eq1} gives
$\|y(t+1)\| \leq \sum_{s=1}^t \|(\overline{W} + \ones e_1^T)^{s+1}\delta(t-s)\|.$
Since $\overline{W}\ones = \textbf{0}$, and $e_1^T \overline{W} = \textbf{0}$, and $\ones e_1^T = (\ones e_1^T)^s$ for any $s=1,2,\ldots$, an inductive argument shows that $W^s = \overline{W}^s + \ones e_1^T$ for any $s=1,2,\ldots$. Thus we can upper bound $\|y(t+1)\|$ using the triangle inequality as follows
$$\|y(t+1)\| \leq \sum_{s=0}^t \left(\|\overline{ W}^{s+1}\delta(t-s)\| + \|(\ones e_1^T)^{s+1}\delta(t-s)\| \right)$$
It is readily verified that for $s=1,2,\ldots$ there holds $e_1^T \delta(s)= 0$. Thus, the second term on the right hand side above is zero, i.e.,
$\|y(t+1)\| \leq \sum_{s=0}^t \|\overline{W}^{s+1}\delta(t-s)\|.$
As a result we have,
$\|y(t+1)\| \leq \sum_{s=0}^t  \lambda^{s+1}\|\delta(t-s)\|.$

Let $t=t^*+T-1$. Using the bound in \eqref{lemma5_delta_ineq} gives,
\begin{align} \label{lemma4_eq1}
 &\|y(t)\| \leq \sum\limits_{s=0}^{t-1} \lambda^{s+1} (n+1)\epsilon(t-1-s)\\
 & = (n+1)\hspace{-2mm}\sum\limits_{s=0}^{t-t^*-1} \lambda^{s+1} \epsilon(t-1-s) + (n+1)\hspace{-2mm}\sum\limits_{s=t-t^*}^{t-1} \hspace{-1mm}\lambda^{s+1}\epsilon(t-1-s).
\end{align}
\vskip -15pt
Consider the first term on the right hand side (RHS) above. Let $\epsilon_{avg}(t^*,T) := \frac{1}{T} \sum_{s=t^*}^{t^*+T-2} \epsilon(s)$. By assumption, the sequence $\{\epsilon(t)\}_{t=t^*}^{t^*+T-2}$ is decreasing, hence by Chebychev's sum inequality \cite{Hardy_et_al} (p. 43-44), $\sum\limits_{s=0}^{t-t^*-1} \lambda^{s+1} \epsilon(t-1-s) \leq \epsilon_{avg}(t^*,T)\sum\limits_{s=0}^{t-t^*-1} \lambda^{s+1} \leq \epsilon_{avg}(t^*,T)\frac{1}{1-\lambda}$, where the latter inequality follows by taking the closed form of the geometric sum. Furthermore, by assumption we have $\sum_{s=t^*}^{t^*+T-2} \epsilon(s) \leq 1$, and hence $\epsilon_{avg}(t^*,T) \leq \frac{1}{T}$, which gives that $\sum\limits_{s=0}^{t-t^*-1} \lambda^{s+1} \epsilon(t-s) \leq \frac{1}{T}\frac{1}{1-\lambda}.$

Now consider the second term on the RHS of \eqref{lemma4_eq1}. By Assumption \ref{assumption_err_bound}, we have $\epsilon(t-s)\leq B$ which allows us to bound the second term as $(n+1)\hspace{-1mm} \sum\limits_{s=t-t^*}^{t-1} \hspace{-1mm}\lambda^{s+1}\epsilon(t-1-s)
 \leq (n+1)B\sum\limits_{s=t-t^*}^{t-1} \hspace{-1mm}\lambda^{s+1} = (n+1)B\lambda^{T}\sum\limits_{s=0}^{t^*-1} \lambda^{s} \leq (n+1)B\lambda^{T}(1-\lambda)^{-1},$
where the latter inequality again follows by taking the closed form of the geometric sum.

Substituting these two bounds back into \eqref{lemma4_eq1} we get
\begin{align}
\|y(t)\| & \leq (n+1)\frac{1}{T}\frac{1}{1-\lambda} + (n+1)B\lambda^{T}\frac{1}{1-\lambda}\\
& = (n+1)\frac{1}{1-\lambda}\left( \frac{1}{T} + B\lambda^{T}\right).
\end{align}
Since we chose $t=t^*+T-1$, this concludes the proof.
\end{myproof}


In order to apply Lemma \ref{lemma_distributed_averaging}, one must show that $\sum_{t=t^*}^{t^*+T-2} \epsilon(t) < 1$. Essentially, this condition states that the variation in the node value $x_1(t)$ during the designated time interval remains bounded. This can be easily ensured, for example, if the value of $x_1(t)$ is monotone. This is the content of the following Lemma.
\begin{lemma}\label{lemma_epsilon_variation}
Suppose that $x_1(t) \in [0,1]$ for all $t$.
Suppose also there exist $t^*,T\in \mathbb{N}_+$ such that $\{x_1(t)\}_{t=t^*}^{t^*+T-1}$ is a monotone sequence. Then $\sum_{t=t^*}^{t^*+T-2} \epsilon(t) \leq 1$.
\end{lemma}
\begin{myproof}
Suppose that $\{x_1(t)\}_{t=t^*}^{t^*+T-1}$ is monotone increasing. Then $\epsilon(t) = |x_1(t+1) - x_1(t)| = x_1(t+1) - x_1(t)$. Substituting this into the sum in question gives a telescoping sum $\sum_{t=t^*}^{t^*+T-2} \epsilon(t) = \sum_{t=t^*}^{t^*+T-2} x_1(t+1) - x_1(t) = x_1(t^*+T-1) - x_1(t^*) \leq 1$. The final inequality follows since $0\leq x_1(t)\leq 1$ for all $t$. A similar argument handles the monotone decreasing case.
\end{myproof}

We now prove Lemma \ref{cor_dist_error} of Section \ref{sec_D-FP_analysis}.
\begin{myproof}[ (Lemma \ref{cor_dist_error})]
Let $\epsilon >0$ and let $t^*\in\mathbb{N}_+$ be arbitrary.
Our task is to show that under the update rule \eqref{empirical_distribution_estimate_recursion}, there exists a $T$ such that if starting at time $t^*$ any action is repeated in $\tilde T\geq T$ consecutive stages, then $\|\hat f_{t^*+\tilde T-1}^i - f_{t^*+\tilde T-1}\| < \epsilon$. We will accomplish this by showing that the update rule \eqref{empirical_distribution_estimate_recursion} fits the template of Lemma \ref{lemma_distributed_averaging}.

Fix a player $j\in\mathcal{N}$ and action $a_j\in \mathcal{A}_j$. Let $f_{j,t}(a_j)$ denote the weight that the empirical distribution $f_{j,t}$ places on $a_j$, and similarly, let $\hat f^i_{j,t}(a_j)$ denote the weight that $\hat f^i_{j,t}$ places on $a_j$. For the purpose of applying Lemma \ref{lemma_distributed_averaging}, let $x_j(0) = 0$, let $x_j(t) := f_{j,t}(a_j)$, $t\geq 1$ and for $i=1,\ldots,n$ let $\hat x_i(0) = 0$, and let $\hat x_i(t) := \hat f_{j,t}^i(a_j)$ for $t\geq 1$. Note that Assumption \ref{assumption_initial_error} is satisfied since $\hat x_i(0) = 0 = x_j(0)$ for all $i$. By \eqref{empirical_distribution_estimate_recursion} and the initialization condition for Algorithm \ref{algo_dfp_example} for $t\geq 0$ we have $\hat x_i(t+1) = \sum_{k\in\mathcal{N}_i} w_{j,k}^i \left(\hat x_k(t) + (x_j(t+1) - x_j(t))\chi_{\{k=j\}} \right).$
Letting $\hat x(t) = (\hat x_i(t))_{i=1}^n\in\mathbb{R}^{n}$ we may express the update rule in more compact notation as
$$
\hat x(t+1) = W_j\left(\hat x(t) + e_j(x_j(t+1) - x_j(t))\right),
$$
where $W_j =(w_{j,k}^i)_{i,k\in\mathcal{N}}$ is the weight matrix as assumed in Lemma \ref{cor_dist_error} and $e_j$ is the $j$-th canonnical vector in $\mathbb{R}^{|\mathcal{A}_j|}$. Note that, after a permutation of the player ordering (which causes no loss in generality), this fits the format of \eqref{update_rule1}. Note also that Assumption \ref{assumption_err_bound} is satisfied since $x_j(t) \in [0,1]$ for all $t$. Furthermore, Assumption \ref{assumption_initial_error} is satisfied since, by construction, $\hat x_i(0)=x_j(0)=0$ for all $i$, and Assumption \ref{assumption_W_matrix} is satisfied by the hypothesis of Lemma \ref{cor_dist_error}.

Now, let $\epsilon(t):=|x_j(t+1) - x_j(t)|$ and suppose that starting at time $t^*$ some action $a^*=(a_1^*,\ldots,a_n^*)\in\mathcal{A}$ is repeated in $T$ consecutive stages, where $T\in\mathbb{N}_+$ is arbitrary. Two cases must be considered---the case that $a_j = a_j^*$ (i.e., the action which defines $x_j$ is in fact the action being repeated by player $j$), and the case that $a_j \not= a_j^*$ (i.e., the action which defines $x_j$ is not being played at all by $j$ during the designated time sequence.)
If $a_j = a_j^*$
then $\{x_j(t)\}_{t=t^*}^{t^*+T-1} = \{f_{j,t}(a_j)\}_{t=t^*}^{t^*+T-1}$ increases monotonically towards 1 (this follows from \eqref{empirical_distribution_recursion_fading}).
Otherwise, if $a_j \not= a_j^*$ then $\{x_j(t)\}_{t=t^*}^{t^*+T-1} = \{f_{j,t}(a_j)\}_{t=t^*}^{t^*+T-1}$ decreases monotonically towards 0.
Since, in either case the sequence is monotone, we have by Lemma \ref{lemma_epsilon_variation} that $\sum_{t=t^*+1}^{t^*+T-2} \epsilon(t) \leq 1$.

Note also that if some action $a^*$ is repeated from time $t^*$ to $t^*+T-1$, then the difference sequence $\{\epsilon(t)\}_{t=t^*}^{t^*+T-2}=\{|f_{j,t+1}(a_j) - f_{j,t}(a_j)|\}_{t=t^*}^{t^*+T-2}$ is decreasing. This follows from \eqref{empirical_distribution_recursion_fading}.

We are now in a position to apply Lemma \ref{lemma_distributed_averaging}.
By the equivalence of finite dimensional norms, there exist constants $c_1$ and $c_\infty$ such that $\|\cdot\| \leq c_1\|\cdot\|_1$ and $c_\infty \|\cdot\|_{\infty}\leq \|\cdot\|$.
Given $j\in \mathcal{N}$ and some action $a_j\in\mathcal{A}_j$ we may choose a constant $T_{a_j}\in\mathbb{N}_+$ sufficiently large such that
$\frac{n+1}{1-\lambda}\left(\frac{1}{T_{a_j}} + B\lambda^{T_{a_j}}\right) < c_\infty\frac{\epsilon}{c_1\sum_{i=1}^{n}|\mathcal{A}_i|}$.
Applying Lemma \ref{lemma_distributed_averaging} we get that if any action $a^*$ is repeated in $T\geq T_{a_j}$ consecutive stages starting at any time $t^*$ then 
\begin{align} \label{lemma5_eq1}
&\max_{i\in \mathcal{N}} |f_{j,t^*+T-1}^i(a_j) - f_{j,t^*+T-1}(a_j)| \\
& = \|(f_{j,t^*+T-1}^i(a_j))_{i=1}^n - f_{j,t^*+T-1}(a_j)\ones\|_{\infty}\\
& = \|\hat x(t^*+T-1) - x_j(t^*+T-1)\ones\|_{\infty} \\
& <\frac{1}{c_\infty} \|\hat x(t^*+T-1) - x_j(t^*+T-1)\ones\| \leq \frac{\epsilon}{c_1\sum_{i=1}^{n}|\mathcal{A}_i|}.
\end{align}
Let $T := \max_{j\in \mathcal{N},a_j\in \mathcal{A}_j} T_{a_j}$. By \eqref{lemma5_eq1} we have $|f_{j,t^*+\tilde T}^i(a_j) - f_{j,t^*+\tilde T}(a_j)| < \frac{\epsilon}{c_1|\sum_{i=1}^{n}|\mathcal{A}_i|}$
for all $\tilde T\geq T$ for all $i,j\in \mathcal{N}$ and for all $a_j\in\mathcal{A}_j$.

Now, fix any player $i\in\mathcal{N}$. Observe that for any $\tilde T\geq T$ we have $\|\hat f^i_{t^*+\tilde T-1} - f_{t^*+\tilde T-1}\| \leq c_1\|\hat f^i_{t^*+\tilde T-1} - f_{t^*+\tilde T-1}\|_1 = c_1\sum_{j\in\mathcal{N}}\sum_{a_j \in \mathcal{A}_j} |\hat f^i_{j,t^*+\tilde T-1}(a_j) - f_{j,t^*+\tilde T-1}(a_j)| \leq c_1 ( \sum_{j=1}^{n}|\mathcal{A}_j|) \frac{\epsilon}{c_1\sum_{j=1}^{n}|\mathcal{A}_j|} = \epsilon$, which is the desired result.
\end{myproof}


}

\bibliographystyle{IEEEtran}
\bibliography{bibliography}

\end{document}